\documentclass[floats,aps,twocolumn,showpacs]{revtex4}
\usepackage{dcolumn}

\usepackage{epsfig}
\newcommand {\la} {\langle}
\newcommand {\ra} {\rangle}
\newcommand {\beq} {\begin{eqnarray}}
\newcommand {\eeqn} [1] {\label{#1} \end{eqnarray}}%
\newcommand {\eol} {\nonumber \\}
\newcommand {\ve} [1] {\mbox{\boldmath $#1$}}

\begin{document}
%
%

\title{
Relation between widths of proton resonances and neutron asymptotic
 normalization coefficients in mirror states of light nuclei 
in a microscopic cluster model.}

\author{
N.\ K.\ Timofeyuk$^{1)}$ and P. Descouvemont$^{2)}$
}                 

\affiliation{
$^{1)}$ Department of Physics, School of Electronics and Physical Sciences, 
University of Surrey, Guildford,
Surrey GU2 7XH, England, UK\\
$^{2)}$ Physique Nucl\'{e}aire Th\'{e}orique et Physique Math\'{e}matique, 
CP229\\
Universit\'{e} Libre de Bruxelles, B1050 Brussels, Belgium
}

\date{\today}

\begin{abstract}
It has been suggested  recently ({\it Phys. Rev. Lett.} 91, 232501 (2003))
that  the widths of  narrow proton resonances  are related to
neutron 
 Asymptotic Normalization Coefficients (ANCs)
of their bound mirror analogs because of 
charge symmetry of nucleon-nucleon interactions.
 This relation is approximated by a simple analytical
formula which  involves  
 proton resonance energies,
 neutron separation energies, charges of residual nuclei and
the range of their strong interaction with the  last nucleon.  
In the present paper, we perform microscopic-cluster model calculations for
the ratio of proton widths to neutron ANCs squared in mirror states 
for several light nuclei. We
compare them to predictions of the analytical formula and to
estimates made within a single-particle potential model.
A knowledge of this ratio can be used to predict  unknown proton
widths for very narrow low-lying resonances in the neutron-deficient
region of the $sd$- and $pf$-shells, which is important for
understanding the nucleosynthesis in the $rp$-process.
\end{abstract}

\maketitle

\section{Introduction}

Many nuclear reactions occur due to virtual  and  real decays
of nuclear levels. The amplitudes of these decays
are important structural characteristics of
nuclei  and their knowledge is needed to predict
reaction cross sections correctly. 
The decay amplitudes are   related in a simple way
to the asymptotic normalization coefficients (ANCs) \cite{BBD77,MTr}.  
The latter
determine the  magnitude of the large distance behaviour of the
projections of nuclear  wave functions   into their decays channels.
A growing interest of the nuclear physics community to the ANCs 
is connected with their applications in nuclear astrophysics.

It has been realised recently  that the amplitudes
of one-nucleon decays of two mirror nuclear states into mirror-conjugated 
channels
should be related if charge symmetry of NN interactions is valid \cite{Tim03}.
As a  consequence,  the ANCs of a pair of particle-bound mirror states  
can be linked by an approximate analytical expression, given
in Ref. \cite{Tim03}, which contains only
nucleon separation energies, charges of the product nuclei and the range 
of the strong
interaction between the last  nucleon and a core. This link  can
be used to predict cross sections of non-resonant proton capture  
if mirror neutron ANCs are known.

In bound-unbound mirror pairs, mirror symmetry of one-nucleon decay
amplitudes manifests itself via a link
between the  width $\Gamma_p$ of
a proton resonance and the   ANC $C_n$
of its mirror bound analog   \cite{Tim03}.
With several
assumptions, one of which suggests that  mirror nuclei have
exactly the same wave functions  
 in the nuclear interior, the ratio
\beq
{\cal R}_{\Gamma} = \Gamma_p/C_n^2,
\eeqn{rg}
can be approximated by the
expression 
\beq
{\cal R}_{\Gamma} \approx {\cal R}_0^{res} = 
\frac{\hbar^2 \kappa_p}{\mu}
\left|\frac{F_l(\kappa_pR_N)}{\kappa_pR_N\,j_l(i\kappa_nR_N)}\right|^2
\eeqn{rnres}
from Ref.   \cite{Tim03}. In this expression,
$l$ is the orbital momentum,
$\kappa_{p(n)} = (2\mu/\hbar^2  \epsilon_{p(n)})^{1/2}$,
$\epsilon_{p(n)}$ is the energy of the proton resonance (neutron separation 
energy), $\mu$ is the reduced mass for the last nucleon plus
core, 
$F_l$ is the regular Coulomb waves function, $j_l$ is
the spherical Bessel function and $R_N$ is the
range of the strong interaction between the last nucleon and the core.
Several examples considered in Ref. \cite{Tim03} have shown  that for narrow 
$l\neq 0$
resonances the predictions of formula (\ref{rnres}) are close to predictions
of a single-particle potential model in which charge symmetry of mirror 
potential wells
and equality of mirror spectroscopic factors are assumed. However, 
for the broad $s$-wave resonance $^{13}$C$(\frac{1}{2}^+)$ these predictions
diverged by about 40$\%$. Moreover, for two mirror pairs
the ratio ${\cal R}_{\Gamma}^{exp}$, constructed  using  
measured proton widths and neutron ANCs,    
deviate   from the predictions
of both the analytical formula and the single-particle model. 
Therefore, an improved   theoretical understanding of the relation
between  widths of proton  resonances and ANCs of their mirror analogs
is required.

A proper understanding of the link between the width of a proton resonance
and the neutron ANC of its miror analog
can be important  for predicting  the rate 
for a particular class of resonant proton capture  reactions at
stellar energies. This class includes reactions that proceed
 via very narrow isolated  resonance states
the proton width $\Gamma_p$ of which
 is either comparable to or much less than its $\gamma$-decay
  width $\Gamma_{\gamma}$. The capture rates for these reactions,
determined by $\Gamma_p\Gamma_{\gamma}/(\Gamma_p+\Gamma_{\gamma})$,
depend strongly on $\Gamma_p$.
 Such narrow resonances can be found in the 
neutron-deficient region of the $sd$ and $pf$ shells (for example, some
levels in $^{25}$Si, $^{27}$P,
$^{33}$Ar, $^{36}$K and $^{43,46}$V) and their study  
is important for understanding  nucleosynthesis in the $rp$ process.
For the  resonances mentioned above $\Gamma_p$ can be much less than 1 eV. 
Direct measurements
of such tiny widths using proton elastic scattering
are impossible. Proton transfer reactions can be  used instead. 
Their analysis (for example, within the
distorted-wave formalism)
provide spectroscopic factors which are combined together with  
single-particle widths to get   necessary partial proton widths 
$\Gamma_p$.
However,  uncertainties in $\Gamma_p$ extracted
using such a procedure  are about 50$\%$ \cite{Ili96}.
These uncertainties arise because of  problems in the 
theoretical treatment of stripping reactions to the continuum and 
due to uncertainties in prediction of   single-particle proton widths.

We suggest an alternative way to determine very small proton widths using
the link ${\cal R}_{\Gamma}$
to the neutron ANCs of their particle-stable mirror analogs. The neutron ANCs
$C_n$ can be determined from experiments with transfer
reactions to bound states, the
theoretical analysis of which encounters less problems than that of 
stripping to continuum. The neutron ANCs can be determined with typical 
accuracy of 10 to 20$\%$. If the uncertainties  in ${\cal R}_{\Gamma}$ are
 less than 10$\%$, then the accuracy of determination of
${\Gamma}_p =  {\cal R}_{\Gamma} C_n^2$  can be between 10 to 30$\%$.
This is more accurate than the distorted wave analysis 
of stripping to continuum can provide. If the determination of $C_n$ requires
experiments with stable beams rather than with radioactive beams, then 
even better accuracy may be achieved.

In this paper, we calculate the ratio ${\cal R}_{\Gamma}$ 
for some resonances in $^8$B, $^{12,13}$N, $^{23}$Al and $^{27}$P 
within a microscopic
cluster model (MCM) which is ideally suited for studying 
decay properties of nuclear levels. 
In our previous work \cite{Tim05}, we  used the same model to study  
 mirror symmetry in
ANCs for bound states of these nuclei. Our study has confirmed 
the
general trend predicted by the simple analytical formula 
of Ref. \cite{Tim05}
for bound-bound mirror pairs which is similar to Eq. (\ref{rnres}).
The deviations from this formula were in general less than 7$\%$
but could increase up to 12$\%$
for loosely bound $s$-states
with a node and for nuclei with strongly excited cores. 
Here, we compare the calculated ratio ${\cal R}_{\Gamma}$ 
to the analytical formula
(\ref{rnres}) and to estimates obtained within a single-particle 
potential  model on the assumption that
single-particle potential wells for mirror states are the same.
First of all, we clarify in Sec. II the meaning of $\Gamma_p$
in formula (\ref{rnres}). Then in Sec. III we compare the predictions of this
formula with exact two-body calculations. 
In Sec.IV
 we explain briefly our microscopic
cluster model. In Sec.  V we study the
ratio ${\cal R}_{\Gamma}$  in the MCM.
 In Sec. VI we discuss mirror symmetry in spectroscopic
factors in bound-unbound mirror pairs and in Sec. VII we summarise
our study and draw conclusions.

\section{Link between $\Gamma_p$ and mirror ANCs}

The approximation (\ref{rnres}) for ${\cal R}_{\Gamma}$
has been derived in Ref. \cite{Tim03} on the
assumption that $\Gamma_p = (\hbar^2\kappa_p/\mu) S_p b_p^2$ \cite{MTr}, 
where $S_p$ is
the spectroscopic factor, $b_p$ is the single-particle ANC of the
Gamow function describing the proton motion in the resonance state, 
and that $S_l^{1/2}b_p$ can be represented by an integral
containing the wave functions of nuclei $A$ and $A-1$ and the interaction
potential between the proton and $A-1$. In reality,  different
definitions for   resonance widths exist \cite{Baz}. These definitions give
similar widths  for narrow resonances but may diverge  
for broad resonances. Therefore, before studying relations
between the proton resonance widths and the ANCs of their
mirror bound analogs it is important to clarify what do we mean
here by a width and what kind of a width enters in Eq.
(\ref{rnres}). We do this in the context of the microscopic R-matrix method 
that we further use in our numerical calculations.

\subsection{Resonance widths in the R-matrix approach}

Let $\Psi_A^{(+)}(\ve{k},\{\ve{r}_i\})$ be 
the wave function of nucleus $A$ above the 
$(A-1)$ + p decay threshold
with the relative momentum $\ve{k}$ in the
decay channel. This wave function   satisfies the Schr\"odinger equation
\beq
H_A \Psi_A^{(+)}(\ve{k},\{\ve{r}_i\}) = E_A \Psi_A^{(+)}(\ve{k},\{\ve{r}_i\}),
\eeqn{SE1} 
where $E_A$ is the total energy of nucleus $A$. 
Let $ \Psi_{A-1}(\{\ve{r}_i\}) $ 
be the wave function   of a bound
state of nucleus $A-1$ with the total energy $E_{A-1}$:
\beq
H_{A-1} \Psi_{A-1}(\{\ve{r}_i\})= E_{A-1}\Psi_{A-1}(\{\ve{r}_i\})
\eeqn{SE2}
Multiplying Eq. (\ref{SE1}) by $\Psi_{A-1}(\{\ve{r}_i\})$, Eq. (\ref{SE2})
by $\Psi_A^{(+)}(\ve{k},\{\ve{r}_i\}) $, subtracting them from each other
and integrating over coordinates of nucleus $A-1$, 
we get
\beq
(T_{rel} + V_0^{coul}(r) - E)
\la \Psi_{A-1}(\{\ve{r}_i\})| \Psi_A^{(+)}(\ve{k},\{\ve{r}_i\}) \ra 
\eol
=
\la \Psi_{A-1}(\{\ve{r}_i\})|\sum_{i=1}^{A-1}V_{iA} - V_0^{coul}(r)|
 \Psi_A^{(+)}(\ve{k},\{\ve{r}_i\}) \ra
\eeqn{ie}
where
 $E = E_A - E_{A-1}$ and
$T_{rel}$ is the kinetic energy operator for the relative motion
between p and $A-1$, 
$V_{ij}$ is the nucleon-nucleon (NN) potential and
\beq
V_0^{Coul}(r) = \frac{Z_{A-1}e^2}{r}
\eeqn{v0c}
is the Coulomb attraction between two point charges
located at distance $r$.
Using the partial wave decomposition 
of the overlap integral in the left-hand side
of Eq. (\ref{ie}),
\beq
\la \Psi_{A-1}(\{\ve{r}_i\})|
\Psi_A^{(+)}(\ve{k},\{\ve{r}_i\}) \ra =
\sum_{lmSM_S M_{J_{A-1}} \sigma} 4\pi i^l 
\eol
\times(J_{A-1}M_{J_{A-1}} \frac{1}{2} \sigma |S M_S) (lm SM_S | J_A M_{J_A})
\eol
\times  \phi_{lS}(k,r) Y_{lm}(\hat{r}) Y^*_{lm}(\hat{k})\, 
\chi_{\frac{1}{2}\sigma}(A),
\eeqn{partial expansion fi}
where $J_i(M_{J_i})$ is the spin (its projection) of nucleus $i$, $l$ is
the orbital momentum and $S$ is the channel spin,  
and expanding the source term (right-hand side) of the inhomogeneous equation 
(\ref{ie}), 
\beq
\la \Psi_{A-1}(\{\ve{r}_i\})| \sum_{i=1}^{A-1}V_{iA} - V_0^{coul}(r)|
\Psi_A^{(+)}(\ve{k},\{\ve{r}_i\}) \ra 
\eol
=
\sum_{lmSM_S M_{J_{A-1}} \sigma} 4\pi i^l 
(J_{A-1}M_{J_{A-1}} \frac{1}{2} \sigma |S M_S) 
\eol
\times (lm SM_S | J_A M_{J_A})
{\cal S}_{lS}(k,r) Y_{lm}(\hat{r}) Y^*_{lm}(\hat{k})\, 
\chi_{\frac{1}{2}\sigma}(A),
\eeqn{partial expansion S}
where $k = \sqrt{2\mu E/\hbar^2}$ and $\mu$ is the reduced mass of proton
and $A-1$,
we get
an inhomogeneous radial equation for the overlap function 
$\phi_{lS}(k,r)$:
\beq
(T_l + V_0^{Coul}(r) - E) \phi_{lS}(k,r) = -S_{lS}(k,r),
\eeqn{inhe2}
where $T_l$  is the kinetic energy operator in the $l$'th partial wave.
This equation can be solved using the Green's functions technique.
In the single channel limit, the regular at  the origin
solution of   Eq. (\ref{inhe2}) reads  
\beq
\phi_{lS}(k,r) = \frac{F_l(\kappa r)}{\kappa r v^{1/2}} - \int_0^{\infty} dr'
\, r'^2 \frac{G_l(r,r')}{rr'} {\cal S}_{lS}(k,r).
\eeqn{phi}
Here $F_l$ is the regular Coulomb wave function,
$G_l(r,r')$ is the outgoing
Green's function for the point charge in the $l$'th
partial wave and $v = \hbar k/\mu$ is velocity.
In Eq. (\ref{phi}), the factor $v^{-1/2}$ is introduced to provide unity flux. 

In the vicinity of
an isolated narrow resonance 
the channel wave function $\phi_{lS}(k,r)$ behaves as  follows \cite{BSA}
\beq
\phi_{lS}(k,r) \approx \frac{\sqrt{\hbar \Gamma_{lS}^0}}{2\kappa}
\frac{
\phi_{lS}^{BSA}(r)}{E - E_R - \frac{i}{2}\Gamma_{lS}^0}
.
\eeqn{BSA}
Eq. (\ref{BSA}), called a bound state approximation,  contains a
square-integrable function $\phi_{lS}^{BSA}(r)$ which is 
defined within some channel radius $a$ taken  outside the range
of the nucleon-nucleus interaction. This function
has the dimension of a bound state wave function. A similar 
bound state approximation
can be also written for the source term ${\cal S}_{lS}(k,r)$:
\beq
{\cal S}_{lS}(k,r) \approx \frac{\sqrt{\hbar \Gamma_{lS}^0}}{2\kappa}
\frac{{\cal S}_{lS}^{BSA}(r)}{E - E_R - \frac{i}{2}\Gamma_{lS}^0}
.
\eeqn{SBSA}
In Eqs. (\ref{BSA}) and (\ref{SBSA}), $E_R$ is the (real) energy 
of the resonance and 
the width $\Gamma_{lS}^0$ is trivially related to
the residue $\gamma_{lS}^2$ in the R-matrix pole:
\beq
\Gamma_{lS}^0 = 2\kappa a \, \gamma_{lS}^2/|O_l(ka)|^2.
\eeqn{gg}

Substituting (\ref{SBSA}) into (\ref{phi}),
using the Green's function from Ref. \cite{green}
\beq
G_l(r,r') = -\frac{2\mu}{\hbar^2\kappa}F_l(\kappa r_<) O_l(\kappa r_>),
\eeqn{GF}
where $O_l = G_l+iF_l$  and $G_l$ is the irregular Coulomb wave function,
and neglecting the term that contains the integral from $r$ to infinity
we can rewrite the wave function (\ref{phi})
at the channel radius $r= a $
 as follows:
\beq
\phi_{lS}(k,a) \approx \frac{i}{2\kappa a v^{1/2}}\left(I_l(\kappa a) - 
U_{lS}O_l(\kappa a)\right) \, ,
\eol
U_{lS} =
1+\frac{2i\sqrt{v\Gamma_{lS}^0}\int_0^a dr r F_l(\kappa r) {\cal S}_{lS}^{BSA}
}{\kappa \hbar^{3/2}(E-E_R-\frac{i}{2}\Gamma_{lS}^0)}
\eeqn{phias}
Here $I_l =  G_l - iF_l$ and $U$ is the collision matrix.
Comparing Eq. (\ref{phias}) with the R-matrix asymptotics of the
wave function $\phi_{lS}$ in the vicinity of the resonance,
\beq
\phi_{lS}(k,a) \approx \frac{i}{2\kappa a v^{1/2}}
\eol \times
\left(I_l(\kappa a) - 
 O_l(\kappa a)\left(1+\frac{i\Gamma_{lS}^0(E)}{E-E_R-\frac{i}{2}
\Gamma_{lS}^0(E)} \right) \right), 
\eeqn{}
we get
\beq
\Gamma_{lS}^0 \equiv \Gamma_{lS}^0(E_R)  = \frac{2\kappa_R}{E_R} 
 \left| \int_0^a
 dr\, r F_l(k_Rr) {\cal S}_{lS}^{BSA}(r) \right|^2,
\eeqn{gamma}
where $k_R = \sqrt{2\mu E_R/\hbar^2}$. The same result can be obtained
in a multi-channel case.
Thus, the partial proton decay width  in the channel $lS$ 
is determined by an integral
that contains the regular Coulomb function of a real argument and a
source term corresponding to the square-integrable wave function
defined in some restricted region.

\subsection{ANCs of bound neutron states}

If the mirror analog of the proton resonance  at energy $E_R$
is a bound state, then the  mirror analog of the
overlap function $\phi_{lS}(k,r) $ is the radial overlap 
integral $I_{lS}(r)$
between the mirror wave functions of nuclei $A$ and $A-1$.
This overlap asymptotically behaves as follows,
\beq
{r}, 
\sqrt{A}\,I_{lS}(r)\approx - C_{lS}\,i^l \kappa_n h^{(1)}_l(\kappa_n r)
,
\,\,\,\,\,\, 
r\rightarrow\infty,
\eeqn{anc}
where $C_{lS}$ is the ANC, 
$h_l^{(1)}$ is the Hankel function of the first kind
and $\kappa_n$ is determined by the separation energy
$\epsilon_n$ of the mirror neutron.
The ANC squared  $C_{lS}^2 $ 
is given by the following expression \cite{Tim98}
\beq
C_{lS}^2 = \frac{4\mu^2}{\hbar^4\kappa_n^2}  \left| \int_0^{\infty}
 dr\, r^2 \kappa_n j_l(ik_nr) {\cal S}_{lS}(r) \right|^2
\eeqn{anc2}
in which  $j_l$ is the spherical Bessel function and 
${\cal S}_{lS}(r) $ is the radial part of the source term
defined by the the left-hand side of expression 
Eq. (\ref{partial expansion S})
in which mirror wave functions of nuclei $A$ and $A-1$ are used
and the Coulomb interactions are absent. 

The main contribution to the integrals in Eqs. (\ref{gamma}) 
and to Eq. (\ref{anc2}) 
for $\Gamma_{lS}^0$ and $C_{lS}^2$
comes from some internal region $r \leq R_N < a$.
If charge symmetry of the NN interactions is valid
and  the  Coulomb differences in the mirror wave functions 
of the resonance and the mirror
bound state in this region can be neglected, 
then the reasoning of Ref.\cite{Tim03}
leads to the analytical formula (\ref{rnres}) for the ratio
between $\Gamma_{lS}^0$ and $C_{lS}^2$. This reasoning
suggests to replace the Coulomb interaction in Eq. 
(\ref{partial expansion S})
at $r\leq R_N$ by a constant equal to $E_R + \epsilon_n$
and then to remove this interaction from the source term
of the proton by changing the function $F_l(k_Rr)$
into some modified function. This function
is a regular solution of the Schr\"odinger equation
with the constant Coulomb potential  
and is equal exactly to the spherical Bessel function
$j_l(ik_nr)$ times the normalization coefficient
$F_l(\kappa_pR_N)/\kappa_pR_N\,j_l(i\kappa_nR_N)$. As the result,
both $\Gamma_{lS}^0$ and $C_{lS}^2$ contain exactly the
same integral and their ratio is detemined by the above normalization
coeffient and the kinematic factor $\hbar^2\kappa_p/\mu$.

It is important 
to notice here that $\Gamma_{lS}^0$ defined by Eq. 
(\ref{gamma})
is a residue in the R-matrix pole so that the   analytical
formula (\ref{rnres}) actually links this residue and the ANC of 
the mirror bound
analog.  If the proton resonance is narrow then $\Gamma^0_{lS}$ 
is approximately equal to the width $\Gamma_{lS}$ of the peak 
in the cross sections of the reactions in which this resonance is populated. 
However, these two widths are not the same. In the particular case of elastic
scattering, the width $\Gamma_{lS}\equiv \Gamma_{lS}(E_R)$ that 
detemines the resonant
phase shift $\tan \delta_{lS} =  \Gamma_{lS}(E)/2(E-E_R)$ is related to
$\Gamma_{lS}^0$ as follows \cite{LT58}:
\beq
\Gamma_{lS} =  \Gamma_{lS}^0 \left(1+\gamma_{lS}^2 S_l'\right)^{-1}
\eeqn{gammaobs}
where $S_l =  {\rm Re}(\kappa aO'_l/O_l)$.  
If a link between physically observed
width $\Gamma_{lS}$ and the neutron ANC is considered,
then Eq. (\ref{rnres}) should be modified by
the factor of $( 1+\gamma_{lS}^2S_l')^{-1}$.


\section{Two-body model}


\begin{figure}[t]
\centerline{\psfig{figure=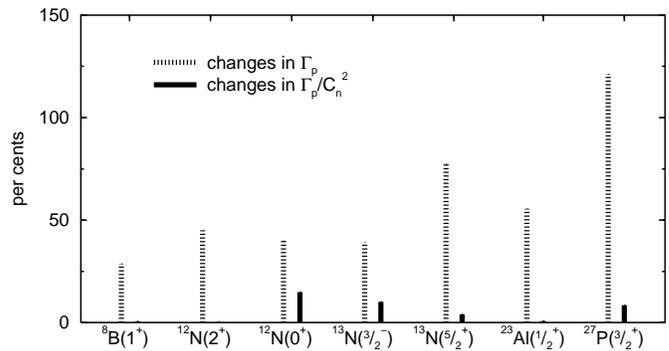,width=0.49\textwidth}
        }
\caption{Changes in proton widths $\Gamma_p$ and in its ratio 
to mirror  neutron ANC squared
$\Gamma_p/C_n^2$   with the choice of two-body nuclear potential well for
a range of nuclei.
}
\end{figure}

According to the analytical formula (\ref{rnres}), 
the ratio between the proton
width and the ANC squared of its  mirror neutron 
is model-independent. It is determined 
only by the proton resonance energy, the neutron separation energy 
and should be the same for any
NN potential employed in calculations. 
We  checked this property for the case of the two-body 
potential model.  
We considered a family of   Woods-Saxon potentials
that give some chosen neutron separation energy $\epsilon_n$, 
and some  chosen  energy of proton resonance $E_p$ when the
Coulomb potential of the uniformly charged sphere was added.
This was achieved by simultaneously varying both 
the depth and the
radius of the Woods-Saxon potential at fixed diffusenesses.
 The actual numerical values of $E_p$  were the same as the
experimental energies of proton resonances in the
lowest states $^8$B($1^+$),  $^{12}$N($2^+$, $0^+$),
$^{13}$N($\frac{3}{2}^-$, $\frac{5}{2}^+$),
$^{23}$Al($\frac{1}{2}^+$) and $^{27}$P($\frac{3}{2}^+$). 
The experimental widths of these resonances are less than 10$\%$ of
their energy so that  the difference between the widths
$\Gamma_{lS}^0$ and $\Gamma_{lS}$  
should be small. In our two-body calculations,
the proton width   (which we refer for simplicity as to 
$\Gamma_p$) has been determined from
transition of the phase shift via 90$^o$. 
As for the mirror neutron separation
energies $\epsilon_n$, they were the same as those obtained in microscopic
calculations below with charge-independent NN interactions.

For different  potentials from the same family, the  
proton widths and mirror neutron   
 ANCs squared   changed
significantly but in such a way that their ratio was roughly
the same.  To illustrate this, we have presented in Fig.1
the changes in proton widths   by thick vertical  dashed lines 
and the changes
in $\Gamma_p/C_n^2$ by the vertical solid lines. While $\Gamma_p$   changes
by 30 to 120$\%$, the changes in $\Gamma_p/C_n^2$ are much smaller.
They are less than 1$\%$  for $^8$B($1^+$), $^{12}$N($2^+$)
and $^{23}$Al($\frac{1}{2}^+$) and 4$\%$ for 
 $^{13}$N($\frac{5}{2}^+$). For $^{27}$P($\frac{3}{2}^+$) 
 the changes in $\Gamma_p/C_n^2$ are 8.5$\%$ as compared to the
121$\%$ changes in $\Gamma_p/C_n^2$. 
Two other cases, $^{12}$N($0^+$) and $^{13}$N($\frac{3}{2}^-$),
are special. Experimentally, these states are seen as narrow resonances with
$\Gamma_p/E_p \approx 0.04$. However, they are so narrow only because of
their very small single-particle strengths, or spectroscopic factors.
In the two-body potential model, the widths of these resonances are
about 30 to 50$\%$ of the resonance energy, depending on the
choice of the two-body potential well. For such broad resonances the
definition of the resonance width (\ref{gamma}) is not precise anymore
and  the ratio $\Gamma_p/C_n^2$  may behave in a different way.
In particular, this ratio is more dependent on the nuclear potential well,
15$\%$ for  $^{12}$N($0^+$) and 10$\%$ for $^{13}$N($\frac{3}{2}^-$),
than in the case of the other narrow resonances.

In Fig. 2 we compare  the ratio $\Gamma_p/C_n^2$ calculated in the 
two-body model with the analytical estimate
${\cal R}_0^{res}$ given by Eq. (\ref{rnres}). 
One can see that   ${\cal R}_0^{res}$ reproduces very well
the general trend in $\Gamma_p/C_n^2$ even for the resonance states
whose widths are not narrow. The difference between $\Gamma_p/C_n^2$
and ${\cal R}_0^{res}$ 
does not exceed 30$\%$, being the smallest for the $d$-wave resonances
(less than 3$\%$) and the largest for the wide two-body resonance
$^{12}$N($0^+$) (about 30$\%$).

The weak sensitivity of the ratio of mirror ANCs to the nuclear
potentials suggests an alternative empirical way to determine this
ratio.  If we assume that mirror neutron and proton single-particle
wells are exactly the same and that the proton
spectroscopic factor $S_p$ can be defined and is equal to
the neutron spectroscopic factor
and $S_n$, then the ratio
${\cal R}_{\Gamma}$ can be approximated by the single-particle ratio
${\cal R}_{\Gamma}^{s.p.}$
\beq
{\cal R}_{\Gamma} \approx {\cal R}_{\Gamma}^{s.p.} \equiv \Gamma_p^{c.s.}/
(b_n^{c.s.})^2,
\eeqn{rsp} 
where the proton width $\Gamma_p^{c.s.}$ and the single-particle ANCs
  $b_n^{c.s.}$ are calculated numerically
for exactly the same nuclear potential well. Further below, we compare
the results of our microscopic calculations for ${\cal R}_{\Gamma}$
with the single-particle estimate ${\cal R}_{\Gamma}^{s.p.}$.
Such a comparison is useful because
unlike   ${\cal R}_0^{res}$, ${\cal R}_{\Gamma}^{s.p.}$ takes into account
the differences in internal wave functions
of mirror nuclei due to the Coulomb interactions.

\begin{figure}[t]
\centerline{\psfig{figure=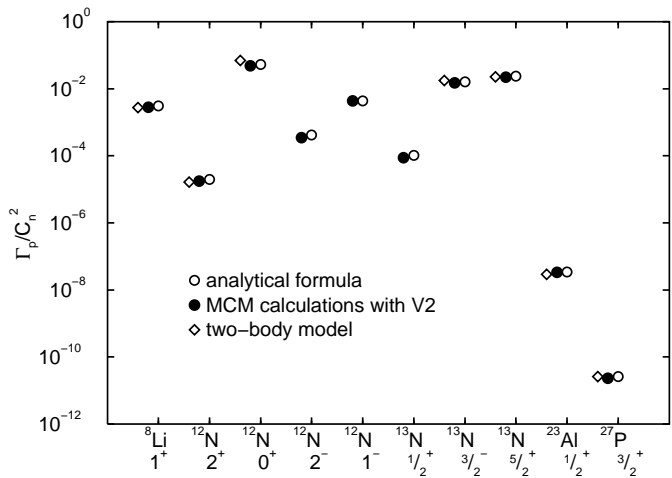,width=0.49\textwidth}
       }
\caption{Ratio of proton width to mirror neutron ANC squared
$\Gamma^0_p/C_n^2$ (given in   units of $\hbar c$)
calculated in the two-body  potential model,
microscopic cluster model  and using the analytical 
formula (\ref{rnres}) for a range of nuclei.
}
\end{figure}


\section{Wave functions in a microscopic cluster model}


To improve our understanding of  ${\cal R}_{\Gamma}$
we use the microscopic cluster model. The MCM 
takes into account the differences in the internal structure of mirror 
nuclei due to the Coulomb interaction and the  core
excitations effects, which were ignored in the derivation of
the analytical formula (\ref{rnres}). On the other hand, it also 
accounts for many-body effects and 
effects caused by non-diagonal Coulomb couplings, which
are absent in the single-particle approximation (\ref{rsp})

The multi-channel cluster wave function for a nucleus $A$ consisting of 
a core $A-1$ and a nucleon $N$ can be represented as follows:
\beq
\Psi^{J_AM_A} = \sum_{lSJ_{A-1} \nu} 
{\cal A} [  \chi_{\frac{1}{2}\tau} [  g_{\nu lS,\omega }^{J_{A-1}} (\ve{r}) 
\otimes 
[\Psi^{J_{A-1}}_{\nu}\otimes  \chi_{\frac{1}{2}}
 ]_S]_{J_A M_A} ] 
 \eol
\eeqn{9}
where ${\cal A} = A^{-\frac{1}{2}}(1-\sum_{i=1}^{A-1} P_{i,A})$ and
the operator $P_{i,A}$ permutes spatial and spin-isospin coordinates
of the $i$-th and $A$-th nucleons.
In this work,
$\Psi^{J_{A-1} }_{\nu}$ is a wave function of nucleus $A-1$ 
with the angular momentum $J_{A-1}$ defined 
either in  translation-invariant harmonic-oscillator shell model, or 
in a multicluster model. 
The quantum number $\nu$ labels
states with the same angular momentum $J_{A-1}$, $S$ is the channel spin
and $\omega $ stands for all the quantum numbers that characterise
the entrance channel. The function $g_{\nu lS,\omega }^{J_{A-1}}$ also
depends on $J_A$ but we omit this index for simplicity.
  
The relative wave function $g_{\nu lS,\omega }^{J_{A-1}} (\ve{r}) = 
g_{\nu lS,\omega }^{J_{A-1}}(r) \,Y_{lm}(\hat{r})$  is determined using the
microscopic R-matrix method, as explained in detail in Ref.
\cite{Des90}.  In this method, the Bloch-Shr\"odinger equation 
is solved for the wave function $\Psi^{J_AM_A}$, which
allows the correct asymptotic behaviour for the relative wave function
$g_{\nu lS,\omega }^{J_{A-1}}$ to be obtained. 
For particle-unstable states
\beq
g_{\alpha,\omega }(r) \approx  A_{\omega}
\frac{\delta_{\omega \alpha} I_{l}(\kappa_{\nu}r) -
U_{\omega \alpha}
O_{l}(\kappa_{\nu}r)}{\kappa_{\omega}v_{\nu}^{1/2}},
\eeqn{11}
where $ I_{l}$ and $O_{l}$ are the ingoing and outgoing Coulomb functions
in the channel $\alpha \equiv\{\nu J_{A-1}lS\}$,
$v_{\nu}$ is the velocity in this channel and $U$ is the collision matrix.
The resonance width $\Gamma_{lS}^0$ 
is determined assuming the Breit-Wigner shape of the 
collision matrix near an isolated resonance at the energy $E_R$
and the width $\Gamma_{lS}$ is determined from $\Gamma_{lS}^0$ 
using Eq. (\ref{gammaobs}).
As for bound states, the asymptotics of the relative
function $g_{\nu lS,\omega }^{J_{A-1}}(r) $ is
\beq
g_{\nu lS,\omega }^{J_{A-1}}(r) \approx C_{ lS,\omega }^{J_{A-1}}
\frac{W_{-\eta_{\nu},l+1/2}(2\kappa_{\nu}r)}{r}
\eeqn{asg}
where $C_{ lS,\omega }^{J_{A-1}}$ is the ANC and
$W$ is the Whittaker function.


\section{Relation between $\Gamma_p$ and mirror ANCs in the MCM}


\begin{table*}
\caption{ Cluster model for nuclei from the first column. The same model
but with mirror-conjugated clusters is used for mirror
analogs of these nuclei. For $^{13}$N, $2c$ and $4c$ stand for
two- and four-cluster model respectively.
 } 
\begin {center}
\begin{tabular}{p{1.8 cm} p{5 cm} p{8 cm} p{2 cm} }
\hline 
\\
 Nucleus & Clustering & Core excitations & References  \\  
\hline
$^8$B & ($\alpha+^3{\rm He})+p$ and  ($\alpha+p)+^3$He &
$^7$Be($\frac{1}{2}^-_{ 1,2}$, $\frac{3}{2}^-_{ 1,2,3,4}$,
$\frac{5}{2}^-_{ 1,2}$, $\frac{7}{2}^-_{ 1,2}$ ) 
and $^5$Li($\frac{3}{2}^-$,$\frac{1}{2}^-$) & 
\cite{Des04}\\
$^{12}$N & $^{11}$C + p  &
$^{11}$C($\frac{1}{2}^-_{ 1,2,3}$, $\frac{3}{2}^-_{ 1,2,3,4,5,6}$,
$\frac{5}{2}^-_{ 1,2,3}$, $\frac{7}{2}^-_{ 1}$)  & \cite{Des99}\\
$^{13}$N$^{2c}$ & $^{12}$C + p  &
$^{12}$C($0^+_{ 1,2,3}$, 
$1^+_{ 1,2,3}$,$2^+_{ 1,2,3,4,5,6}$) & \cite{TBD97}\\
$^{13}$N$^{4c}$ & $\alpha+\alpha+\alpha+p$  &
$^{12}$C($0^+_{ 1,2,3}$,  $2^+_{ 1,2,3}$) & \cite{Duf97}\\
$^{23}$Al & $^{22}$Mg + p  &
$^{22}$Mg($0^+_{ 1}$, $1^+_{ 1}$,
$2^+_{ 1,2,3}$,$3^+_{ 1}$, $4^+_{ 1,2}$) & \cite{Tim05}\\
$^{27}$P & $^{26}$Si + p  &
$^{26}$Si($0^+_{ 1}$, $2^+_{ 1}$, $4^+_{ 1}$) & \cite{Tim05}\\
\hline
\end{tabular}
 
\end{center}
\label{table1}
\end{table*} 

In this section we investigate the ratio ${\cal R}_{\Gamma}$ for
the following $0p$ and $sd$ shell resonances:
$^8$B(1$^+$), $^{12}$N(2$^+$,1$^+$,2$^-$,1$^-$), 
$^{13}$N($\frac{1}{2}^+$,$\frac{5}{2}^+$,$\frac{3}{2}^-$), 
$^{23}$Al($\frac{1}{2}^+$) and $^{27}$P($\frac{3}{2}^+$). The cluster models
we used are listed in Table I together with  references
for further details. The internal
structure of these clusters is represented
by the shell model Slater determinants composed of the
following single-particle oscillator
wave functions: $0s$ for $\alpha$, $^3$H and $^3$He clusters,
$0s$ and $0p$ for the  $^{11}$B and $^{11,12}$C clusters and
$0s$, $0p$ and $0d_{\frac{5}{2}}$ for the $^{22}$Ne, $^{22}$Mg, $^{26}$Mg
and $^{26}$Si clusters. Some excited states of these clusters, appeared
in teh shell model calculations, are taken into account. They are also listed
in Table I.
The products of the proton decay and  
their mirror analogs are always considered to
be in their  ground states. 

We use
well adapted effective NN interactions for such calculations, namely,
 the  Volkov potential V2 \cite{volkov}
and the Minnesota (MN) potential \cite{minnesota}.
The two-body spin-orbit force
\cite{BP81} and the Coulomb interaction are also included.
More details on the conditions of calculations can be found in
Ref. \cite{Tim05} and references therein.

Each of V2 and MN have one adjustable parameter that gives the
strength of the odd NN potentials $V_{11}$ and $V_{33}$. To get meaningful
values of ANCs and proton widths, this
parameter should be fitted in each individual case
to reproduce the experimental neutron
separation energy $\epsilon_n$ or the resonance energy $E_p$
of the proton. 
In most cases, the same choice of the adjustable parameter for mirror states
does not  let to reproduce exactly $\epsilon_n$ and $E_p$.
Therefore, we consider here two cases: ($A$) We keep the same value
of the adjustable parameter for both nuclei of a mirror pair thus imposing
charge symmetry of the NN interactions. The value of this parameter
is fitted to reproduce experimental value of $E_p$.
($B$)
We use slightly different adjustable parameters in mirror nuclei to
reproduce simultaneously $\epsilon_n$ and $E_p$.
This simulates charge symmetry breaking  of the effective NN interactions
which should be a consequence of the charge symmetry breaking in 
realistic NN interactions. For both cases we have calculated
the ratio ${\cal R}_{\Gamma_0}^{MCM} = \Gamma_{lS}^0/C_{lS}^2$ 
that we compare to the analytical estimate ${\cal R}_0^{res}$.
We have also calculated the ratio
${\cal R}_{\Gamma}^{MCM} = \Gamma_{lS}/C_{lS}^2$ which we compare
to the single-particle estimate ${\cal R}_{\Gamma}^{s.p.}$.
In $^8$B(1$^+$) and $^{12}$N(2$^+$), where two different values of the channel
spin $S$ are possible, we use the sums $\sum_{S}\Gamma_{lS}^0$,
$ \sum_{S}\Gamma_{lS}$ and $\sum_{S}C_{lS}^2$  when constructing
these ratios.

\subsection{Calculations with charge-independent NN interactions}

\begin{figure*}[t]
\centerline{\psfig{figure=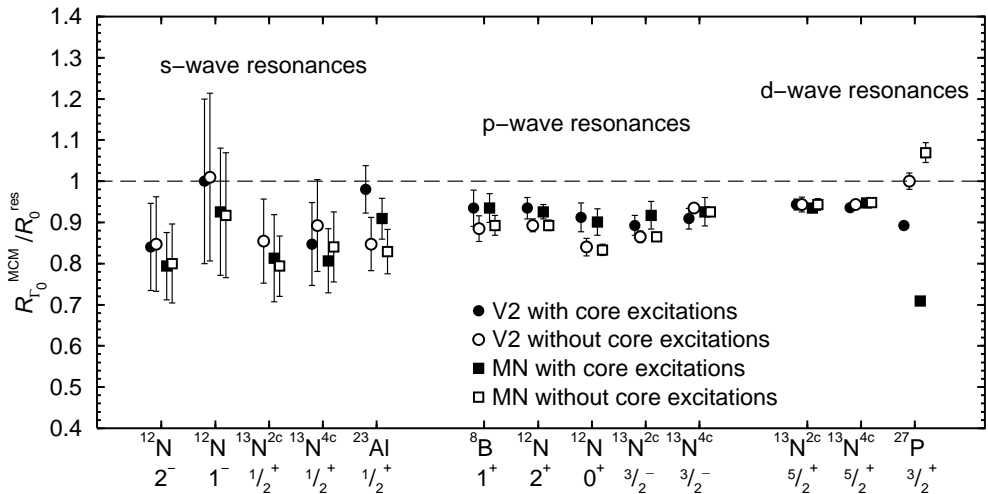,width=0.65\textwidth}
        }
\caption{Ratio between the predictions
${\cal R}_{\Gamma_0}^{MCM}$ from
the microscopic calculations  and the analytical estimate
${\cal R}_{0}^{res}$ of the analytical formula (\ref{rnres}).
The microscopic calculations are performed
for the V2  and MN  potentials
with and without taking core excitations into account. 
Charge symmetry of NN interactions is assumed. 
 Both four-cluster ($4c$) and two-cluster ($2c$) 
calculations for $^{13}$N are shown.
}
\end{figure*}

\begin{table*}
\caption{Asymptotic normalization coefficients  squared
$C_{lj}^2$ (in fm$^{-1}$),  
spectroscopic factors $S_{lj}$, single-particle ANCs squared
 $b_{lj}^2 = C_{lj}^2/S_{lj}$ (in fm$^{-1}$)
and r.m.s. radii $\la r^2_{lj}\ra^{1/2}$ (in fm)
for the nuclei from the first column. 
The calculations have been performed with  two NN potentials, V2 and MN. 
The experimental neutron separation energies have been reproduced. } 
\begin {center}
\begin{tabular}{ p{1.8 cm} p{0.9 cm} p{2.2 cm} p{1.8 cm}  p{2.2  cm} 
p{1.2 cm} p{2.2 cm}
 p{1.8 cm}  p{2.2 cm}  p{1.0 cm}  }
\hline 
\\
 & & \multicolumn{4}{c}{V2} & \multicolumn{4}{c}{MN} \\
 Nucleus & $lj$ & $C_{lj}^2$ & $S_{lj}$ & $b_{lj}^2$ & 
 $\la r^2_{lj}\ra^{1/2}$ & $C_{lj}^2$ & $S_{lj}$ & $b_{lj}^2$ & 
 $\la r^2_{lj}\ra^{1/2}$\\
\hline
 $^8$Li(1$^+$) 
& $p_{1/2}$ & 0.0135 & 0.051& 0.263 &  4.43 & 0.0255 &  0.097 & 0.263 & 4.42\\ 
& $p_{3/2}$ &  0.1378 &  0.514& 0.268 &  4.46  &   0.1261 &  0.525 & 0.240 
& 4.30 \\   
 & total & 0.1513 &  0.566 &  &  &  0.1516 &  0.622 & &\\

$^{12}$B(2$^+$)      
    & $p_{1/2}$ & 0.536 & 0.490 &   1.10 & 3.79  & 0.490 & 0.500 & 0.979 & 
     3.68 \\
      & $p_{3/2}$ & 0.0235 & 0.0177 &  1.33 &  3.99 &  0.0202 &  0.0143 & 
      1.412 &  4.06 \\
     & total & 0.560 &    0.507 &  & &    0.510 &   0.514 &  & \\

$^{12}$B(0$^+$)
& $p_{3/2}$ & 0.0835 &  0.607 & 0.138 &  4.77  & 0.0638 & 0.515 &   0.124 & 
4.57 \\

$^{12}$B(2$^-$)
& $s_{1/2}$ & 2.721 &    0.898 &  3.03 &   4.99  &
 2.504 &   0.934 &   2.68 &    4.78 \\

 $^{12}$B(1$^-$)
& $s_{1/2}$ & 1.292 &   1.016 & 1.27 &   6.49 & 1.148 &   0.992 &  1.158 &   
6.25 \\

   $^{13}$C($\frac{1}{2}^+$)$^{2c}$ &
$s_{1/2}$ &
   3.46$\pm$0.02 & 0.94  & 3.70$\pm$0.02 & 4.98 & 3.41$\pm$0.03 & 1.04 & 
   3.28$\pm$0.02 & 4.80 \\

      $^{13}$C($\frac{1}{2}^+$)$^{4c}$ &
$s_{1/2}$ &
   3.39$\pm$0.02 & 0.94  & 3.60$\pm$0.01 & 4.98 &
 2.77$\pm$0.03 & 0.92 & 3.01$\pm$0.03 & 4.77 \\

  $^{13}$C($\frac{3}{2}^-$)$^{2c}$ & $p_{3/2}$ &  
   0.0992 & 0.271 & 0.365 & 4.21 & 0.0931& 0.283 & 0.331 & 4.09 \\

  $^{13}$C($\frac{3}{2}^-$)$^{4c}$ &
$p_{3/2}$ &  
 0.129$\pm$0.001 & 0.370 & 0.350$\pm$0.002 & 4.18  &
  0.100$\pm$0.002 & 0.363 & 0.274$\pm$0.004 & 3.90\\

   $^{13}$C($\frac{5}{2}^+$)$^{2c}$ &
$d_{5/2}$ &
   0.0272 & 0.881 & 0.031 & 4.04 & 0.0222& 0.873 & 0.025 & 3.86\\

      $^{13}$C($\frac{5}{2}^+$)$^{4c}$ &
$d_{5/2}$ &
   0.0213$\pm$0.0006 & 0.81 & 0.0263$\pm$0.0006 & 3.90 &
 0.0153$\pm$0.0003 & 0.84 & 0.0183$\pm$0.0004 & 3.59\\

   $^{23}$Ne($\frac{1}{2}^+$)$^{a)}$ &
$s_{1/2}$ &
   3.52 & 0.202 & 17.4  & 4.31 & 2.37 & 
     0.154 & 15.4  & 4.19\\
    
   $^{23}$Ne($\frac{1}{2}^+$)$^{b)}$ &
$s_{1/2}$ &
   18.49 & 1.08 & 17.1 & 4.26 & 16.3  & 1.08 &   15.0  & 4.12\\
 
   $^{27}$Mg($\frac{3}{2}^+$)$^{a)}$ &
$d_{3/2}$ &
   1.89$\pm$0.03 & 0.744$\pm$0.002 & 2.54$\pm$0.03  & 3.63 & 0.695$\pm$0.012 & 
     0.318$\pm$0.002 & 2.18$\pm$0.02 & 3.56 \\
    
   $^{27}$Mg($\frac{3}{2}^+$)$^{b)}$ &
$d_{3/2}$ &
   2.61$\pm$0.04 & 1.08 & 2.43$\pm$0.03 & 3.60 & 2.03$\pm$0.04  & 1.08 
   & 1.88$\pm$0.04 &
   3.47 \\
\hline
\multicolumn{10}{l}{$^{2c}$ - two-cluster model} \\
\multicolumn{10}{l}{$^{4c}$ - four-cluster model }\\
\multicolumn{10}{l}{$^{a)}$ - multi-channel cluster model} \\
\multicolumn{10}{l}{$^{b)}$ - single-channel cluster model} \\
\end{tabular}
 
\end{center}
\label{table11}
\end{table*}

The calculated ratio ${\cal R}_{\Gamma_0}^{MCM}$
for the V2 potential is plotted in Fig.2 
together with the
prediction ${\cal R}_{0}^{res}$ of the analytical formula (\ref{rnres})
and with the single-particle estimate ${\cal R}_{\Gamma}^{s.p}$.
As seen in  Fig.2, both ${\cal R}_{0}^{res}$ and
${\cal R}_{\Gamma}^{s.p}$ describe 
very well the general trend in the ${\cal R}_{\Gamma_0}^{MCM}$
behaviour. 
 
To see the differences between  ${\cal R}_{\Gamma_0}^{MCM}$
and ${\cal R}_0^{res}$ we have plotted
the ratio ${\cal R}_{\Gamma_0}^{MCM}/{\cal R}_0^{res}$ in Fig.3. 
The error bars
 in this figure represent
uncertainties in ${\cal R}_0^{res}$ due to the choice of $R_N$.  
The estimation of these errors is similar to that described in Ref.
\cite{Tim05} for bound-bound mirror pairs.

According to Fig.3, 
the effect of different NN potential choices
is normally less than 7$\%$ for the $0p$ shell nuclei, being the smallest for
the $p$-wave resonances (less than 2$\%$) and for the
$d$-wave resonance $^{13}$N($\frac{5}{2}^+$).  Slightly bigger effect,
about 9$\%$, is seen for the $s$-wave resonance $^{12}$N(1$^-$).
For the $sd$ shell nuclei $^{23}$Al and $^{27}$P, the sensitivity on
the NN potential depends on whether the core excitations are taken into
account or not. Without core excitations, the sensitivity to the NN potential
choice is less than 6$\%$, which is similar to the case of the 
$0p$ shell nuclei. When core excitations are present, this 
sensitivity increases up to  10$\%$ for $^{23}$Al and
30$\%$ for $^{27}$P.

Our previous study  of ANCs in bound-bound
mirror pairs \cite{Tim05}  has shown that core excitations 
can be responsible for differences between the MCM calculations
for the ratio of mirror ANCs
 and the predictions of the analytical formula 
(7) of Ref. \cite{Tim03}. In the present paper, 
we check how important the core excitations are
in  bound-unbound mirror states. Fig.3 shows the  
ratio ${\cal R}_{\Gamma_0}^{MCM}/{\cal R}_0^{res}$   calculated  
both in the single-channel (no core excitations)
and the multi-channel
(including the  core excitations from Table I)
cluster model.  
One can see that
for the $0p$ shell nuclei the results obtained with and without taking
core excitations into account differ by no more than 7$\%$.
The influence of the 
core excitations on ${\cal R}_{\Gamma_0}^{MCM}$ 
becomes more important for  nuclei in the middle of the 
$sd$ shell. For $^{23}$Al - $^{23}$Ne, this influence
 is 9 - 12$\%$. A similar effect is present  
in the calculations with the V2 potential
for $^{27}$P($\frac{3}{2}^+$) - $^{27}$Mg($\frac{3}{2}^+$).
However, for  MN, the influence of core excitations
on ${\cal R}_{\Gamma_0}^{MCM}$ is much stronger, about 37$\%$.
For  the MN potential, the $d$-wave
$^{26}$Si(0$^+$) + p configuration  in $^{27}$P($\frac{3}{2}^+$) becomes 
three times weaker than the $s$-wave $^{26}$Si(2$^+$) + p configuration.
In weak configurations, effects of charge  symmetry breaking due to 
the Coulomb
interactions can be more noticeable. In the  case of 
$^{27}$P($\frac{3}{2}^+$) - $^{27}$Mg($\frac{3}{2}^+$), 
the significant difference between
${\cal R}_{\Gamma_0}^{MCM}$ and ${\cal R}_0^{res}$ 
coincides with similar  mirror symmetry breaking in the spectroscopic
factors of the $^{26}$Si(0$^+$) + p and $^{26}$Mg(0$^+$) + n
configurations, which is about 33$\%$ for MN. 
For V2, the
$d$-wave $^{26}$Si(0$^+$) + p configuration dominates 
and the mirror symmetry
breaking for spectroscopic factor of this configuration is only 4$\%$.

The average difference between
${\cal R}_0^{res}$ and ${\cal R}_{\Gamma_0}^{MCM}$ is about 10$\%$. This is
larger than the average deviation between the microscopic calculations
with charge-independent NN interactions
and the predictions of analytical  formula for the ratio of mirror ANCs
squared for bound-bound mirror pairs, obtained  in Ref. \cite{Tim05}.

\subsection{Calculations  with charge breaking symmetry NN interactions}

The calculated ANCs for bound neutron states are shown in 
Table II. We show them in the $lj$ coupling
scheme that is widely accepted in the analysis of transfer reactions,
in which these ANCs can, or have been determined.
Transition from the $lS$ coupling scheme to the $lj$ coupling scheme  
is performed using  standard technique.
For $^8$Li(1$^+$) and $^{12}$N(2$^+$)  
we also show the sum of the ANCs squared 
$C_l^2 = \sum_{j}C_{lj}^2 = \sum_{S}C_{lS}^2$. 
We have also calculated the radial part $I_{lj}(r)$ of the
overlap integral $\la A | A-1\ra$ and the
spectroscopic factors, 
r.m.s. radii $\la r^2_{lj}\ra^{1/2} = 
\left(\int_0^{\infty} dr\, r^4 I^2_{lj}(r)/
\int_0^{\infty} dr\, r^2 I^2_{lj}(r)\right)^{1/2}$
and single-particle ANCs
$b_{lj} = C_{lj}/S_{lj}^{1/2}$ associated with it. We show them  
in Table II as well.

For proton unbound states we have calculated both the
widths $\Gamma_{lS}^0$ and $\Gamma_{lS}$ which  are shown
in Table III together with experimentally measured widths. 
$\Gamma_{lS}^0$ slightly depends on the choice of the channel radius $a$.
We tried to choose this radius to be large enough, between 8.0 and 10.0 fm.
In this region, the sensitivity of $\Gamma_{lS}^0$ to $a$ is about 2$\%$
for narrow resonances but can reach about 7$\%$ for
broad $s$-wave resonances. 
The coefficient
$\alpha = \Gamma_{lS}/ \Gamma_{lS}^0$ that connects 
$\Gamma_{lS}$ and $\Gamma_{lS}^0$ is also $a$-dependent so that
the sensitivity of $\Gamma_{lS}$ to the choice of $a$ decreases.
These coefficients, shown in Table
III, are between 1.01 and 1.05 for narrow resonances
but can increase up to 1.38 for broad $s$-wave resonances.

The averaged over two NN potentials ratios 
${\cal R}_{\Gamma_0}^{MCM}$ and
${\cal R}_{\Gamma}^{MCM}$ are shown
in Table IV together with the
analytical estimate ${\cal R}_{\Gamma_0}^{MCM}$ 
and the single-particle estimate ${\cal R}_{\Gamma}^{s.p.}$.
The ratios ${\cal R}_{\Gamma_0}^{MCM}/{\cal R}_{0}^{res}$
and  ${\cal R}_{\Gamma}^{MCM}/{\cal R}_{\Gamma}^{s.p.}$
are also presented in Fig. 4.

\subsubsection{$^8{\rm B} -  ^8{\rm Li}$}

According to Table II, 
the ANCs in small $j=1/2$ components of the $\la^8$Li$|^7$Li$\ra$
overlap integral strongly depend on the choice of the NN potential.
However, in dominant components $j=3/2$ they are very similar. 
The total neutron ANC squared $C_1^2$ are practically the same for 
both the NN potentials used in calculations. 
The  total widths $\Gamma_l = \sum_S \Gamma_{lS}$
are also practically the same for both V2 and MN (see Table III).
The ratio ${\cal R}_{\Gamma_0}^{MCM}$ varies within 5$\%$ with 
the NN potential choice. Its average
value of $(1.80\pm 0.04)\times 10^{-3}$ is about 10 $\%$ smaller
than the prediction ${\cal R}_0^{res}$ = $(2.06\pm 0.04)\times 10^{-3}$
of the analytical formula 
(\ref{rnres}).
The ratio ${\cal R}_{\Gamma}^{MCM}$ = 
$(1.73\pm 0.03)\times 10^{-3}$
is close to the single-particle estimate 
${\cal R}_{\Gamma}^{s.p.}$ = (1.78$\pm$0.01)$\times 10^{-3}$.

The neutron ANCs for $^8$Li(1$^+$) have been determined  
from experimental study of the
transfer reaction $^{13}$C($^7$Li,$^8$Li)$^{12}$C in
Ref. \cite{Tra03}.
It was found that
$C^2_{1\frac{3}{2}}/C^2_{1\frac{1}{2}}$ = 0.22(3) 
and  $C^2_1$ = 0.082$\pm$0.009 fm.
The measured   ratio $C^2_{1\frac{3}{2}}/C^2_{1\frac{1}{2}}$ is
in excellent agreement with our prediction of 0.20 with the MN force
but is about twice as high as the prediction of
0.098 made with the V2 potential. The total value of the 
experimental ANC squared
is significantly lower than the predictions of the MCM.
 
Two experimental values of the  proton width $\Gamma_l$ 
for the 1$^+$ resonance
are available, $37 \pm 5$ keV from the 
$^7$Be(p,$\gamma$)$^8$B reaction \cite{Aiz88} and $31 \pm 4$ keV 
from elastic scattering $^7$Be + p \cite{Ang03}.
The MCM calculations give larger widths,
50.9 keV and 52.7 keV for  V2 and MN 
respectively. 
The ratio ${\cal R}_{\Gamma}^{exp}$, calculated with $\Gamma_l = 
37 \pm 5$ keV,
is (2.29$\pm$0.28)$\times 10^{-3}\, \hbar c$. This is significantly
larger than the MCM value of  $(1.73\pm 0.03)\times 10^{-3}\, \hbar c$ 
obtained in this work (see Table IV). 
With more recent value, $\Gamma_l = 31 \pm 4$ keV,
this ratio is smaller,
${\cal R}_{\Gamma}^{exp} = (1.92\pm 0.23) \times 10^{-3}\, \hbar c$,
and it agrees with ${\cal R}_{\Gamma}^{MCM}$ = 
$(1.73\pm 0.03)\times 10^{-3}$ within the error bars.

\subsubsection{$^{12}{\rm B} - ^{12}{\rm N}$}

\begin{table*}
\caption{ The proton widths $\Gamma_{lS}^0$,  $\Gamma_{ls}$ 
(in keV) and the  coefficient 
$\alpha = \Gamma_{lS}^0/\Gamma_{lS} = 
1+ \gamma_{lS}^2S_l'$ 
that relates them (see Eq. (\ref{gammaobs}),
for the resonances
from the first column. Also shown are the resonance energies $E_{res}$
(in keV), orbital momenta $l$ and channel radii $a$ (in fm). For the resonances
$^8$B(1$^+$) and $^{12}$N(2$^+$) two values of channel spin $S$ are possible.
The proton widths in  these channels are shown separately.
The calculations have been performed with  two NN potentials, V2 and MN. 
 } 
\begin {center}
\begin{tabular}{ p{1.8 cm} p{0.9 cm} p{0.4 cm} p{1.0 cm}  p{1.0 cm}
p{1.9  cm} p{1.3 cm} p{1.9 cm}
 p{1.9 cm}  p{1.3 cm}  p{1.8 cm} p{1.7 cm} }
\hline 
\\
 & & & & &\multicolumn{3}{c}{V2} & \multicolumn{3}{c}{MN} & \\
 Resonance &  $E_{res}$ & $l$ & $a$ & & $\Gamma_{lS}^0$ & $\alpha$ & 
 $\Gamma_{lS}$ & 
 $\Gamma_{lS}^0$ & $\alpha$ & $\Gamma_{lS}$ & $\Gamma_p^{exp}$\\
\hline
$^8$B(1$^+$) & 633 & 1 & 10.0 & $S$=1 & 42.4 & 1.042 & 40.7 & 49.8 
& 1.050& 47.4 &\\
 & & & & $S$=2 & 10.3 & 1.010  &  10.2 & 5.36 & 1.005 & 5.33  &\\
 & & & & total &  & &    50.9  & & &  52.7 & 37$\pm$5 {\cite{Aiz88}} \\ 
  & & & & & & & & & & & 31$\pm$4 {\cite{Ang03}} \\ 
                                          
$^{12}$N(2$^+$) & 359 & 1 & 8.55 & $S$=1 & 0.949 & 1.032 & 
                               0.920 & & & &\\
& & & & $S$=2 & 0.388 & 1.013 & 
                                          0.383 & & & & \\
 & & & & total &  & &   1.30  & & & & $<20$ {\cite{Aiz90}}     \\ 
& & & 8.75 & $S$=1 & & & & 0.879 & 1.028 & 
                                          0.855 &\\
& & & & $S$=2 & & & & 0.365 & 1.011 & 
                                          0.361 &\\
 & & & & total &  & & & & &  1.22  & $<20$ {\cite{Aiz90}}   \\ 
$^{12}$N(0$^+$) & 1839  & 1 & 8.55 &   &  344 & 1.034 & 
                           332 &  260 & 1.026 &  253 
                          & 68$\pm$21 {\cite{Aiz90}}   \\ 
 $^{12}$N(2$^-$) & 589  & 0 & 8.75 &   &  150 & 1.26  & 
                            119 &  138 & 1.24 & 111  
                           & 118$\pm$4  {\cite{Aiz90}} \\ 
 $^{12}$N(1$^-$) & 1199  & 0 & 8.55 &   &   &    & 
                   &  771 & 1.108 &  696  & 750$\pm$250 {\cite{Aiz90}}  \\
 & & & 8.75 &   &  821  &  1.109  &  740
                                           &   &  &   & \\ 
$^{13}$N($\frac{1}{2}^+$)$^{2c}$ & 422  & 0 & 8.5 &   &   &    & 
                     & 49.6 & 1.33 & 37.3  & 31.7$\pm$0.8 {\cite{Ajz91}} \\
 & & & 9.5 &  &  50.4 & 1.30  & 38.8 & 46.3  &  1.25  & 37.0  &\\
$^{13}$N($\frac{1}{2}^+$)$^{4c}$ & 422  & 0 & 8.0 &   & 51.2 & 1.38  & 37.1 
                            & 43.4 & 1.32 & 32.8 & 31.7$\pm$0.8 {\cite{Ajz91}}\\
 & & & 8.5 &  & 49.3  & 1.33 & 37.1 & 41.6  &  1.28  & 32.6 & \\ 
$^{13}$N($\frac{3}{2}^-$)$^{2c}$ &  1557  & 1 & 8.5 &   & 118 & 1.021  & 
116  & 114 & 1.020 & 112 & 62$\pm$4 {\cite{Ajz91}}\\ 
 & & & 9.5 &  & 116  & 1.015 & 114 & 112  &  1.014  & 110 \\ 
$^{13}$N($\frac{3}{2}^-$)$^{4c}$ &  1557  & 1 & 8.5 &   & 164 & 1.028  & 159  
              & 132 & 1.023 & 129 & 62$\pm$4 {\cite{Ajz91}}\\  
 & & & 8.9 &  & 160  & 1.024 & 156 & 128  &  1.020  & 126 & \\      
$^{13}$N($\frac{5}{2}^+$)$^{2c}$ &  1607  & 2 & 8.5 &   & 69.5 & 1.046  
& 66.4   & 57.1 & 1.038 & 55.0 & 47$\pm$7 {\cite{Ajz91}} \\  
 & & & 9.5 &  & 68.0  & 1.031 & 66.0 & 55.8  &  1.025  & 54.4 & \\  
$^{13}$N($\frac{5}{2}^+$)$^{4c}$ &  1607  & 2 & 8.0 &  & 57.4 & 1.046 & 54.9
     & 41.3 & 1.033 & 40.0 & 47$\pm$7 {\cite{Ajz91}}\\  
 & & & 8.5 &  & 56.0  & 1.037 & 54.0 & 39.6  &  1.026  & 38.6 & \\ 
$^{23}$Al($\frac{1}{2}^+$)$^{a)}$ &  405  & 0 & 9.0 &   &  
2.01$\times$10$^{-2}$ & 1.015   & 
      1.98$\times$10$^{-2}$  & 1.26$\times$10$^{-2}$ & 1.010 
      & 1.25$\times$10$^{-2}$ & \\   
$^{23}$Al($\frac{1}{2}^+$)$^{b)}$ & 405   & 0 & 9.0 &   &  
9.19$\times$10$^{-2}$ & 1.071   & 
      8.58$\times$10$^{-2}$  & 8.19$\times$10$^{-2}$ & 
      1.063 & 7.70$\times$10$^{-2}$ & \\   
$^{27}$P($\frac{3}{2}^+$)$^{a)}$ &  340  & 2 & 9.5 &   &  
7.18$\times$10$^{-6}$ & 1.003   & 
      7.16$\times$10$^{-6}$  & 2.14$\times$10$^{-6}$ & 1.001 & 
      2.14$\times$10$^{-6}$ & \\
$^{27}$P($\frac{3}{2}^+$)$^{b)}$ &  340  & 2 & 9.5 &   &  
1.04$\times$10$^{-5}$ & 1.005   & 
      1.04$\times$10$^{-5}$  & 8.26$\times$10$^{-6}$ & 
      1.004 & 8.23$\times$10$^{-6}$ & \\
\hline
\multicolumn{10}{l}{$^{2c}$ - two-cluster model} \\
\multicolumn{10}{l}{$^{4c}$ - four-cluster model }\\
\multicolumn{10}{l}{$^{a)}$ - multi-channel cluster model} \\
\multicolumn{10}{l}{$^{b)}$ - single-channel cluster model} \\
\end{tabular}
 
\end{center}
\label{table0}
\end{table*}

The   ANCs squared obtained 
for the $2^+$, $2^-$ and $1^-$ states of $^{12}$B
with V2 and MN 
differ by 9-13$\%$ from each other. For
$^{12}$B(0$^+$) this difference is larger, about 30$\%$.
The  neutron spectroscopic factors 
change by 1-6$\%$ with different choices of the NN potential
except for  $0^+$, where this difference reaches 18$\%$.
The proton widths in $^{12}$N 
differ by 6-9$\%$ for the $2^+$, $2^-$ and $1^-$
states and by 32$\%$ for the $0^+$ state. The sensitivity of the
calculated ratios ${\cal R}_{\Gamma_0}^{MCM}$ and ${\cal R}_{\Gamma}^{MCM}$
to the NN potential is less than 3$\%$ except for the broad
resonance 1$^-$ where it reaches 6$\%$.

The average value of ${\cal R}_{\Gamma_0}^{MCM}$  is lower
than the analytical estimate ${\cal R}_{0}^{res}$ by
 15$\%$, 18$\%$, 28$\%$ and 16$\%$ for the 
 $2^+$, $0^+$, $2^-$ and $1^-$ states respectively.
The other ratio, ${\cal R}_{\Gamma}^{MCM}$ compares with the single-particle
estimate ${\cal R}_{\Gamma}^{s.p.}$ in a different way for each state.
For the narrow resonance 2$^+$, they agree within 2$\%$. For the
resonance 0$^+$, which is broad in the single-particle potential model
($\Gamma_l \sim 0.7$ MeV)  but
is much narrower in the MCM ($\Gamma_l \sim 0.3$ MeV), 
${\cal R}_{\Gamma}^{s.p.}$ is larger than the
analytical estimate ${\cal R}_{0}^{res}$ 
(for most other cases it is smaller) and
discrepancy between ${\cal R}_{\Gamma}^{MCM}$ and ${\cal R}_{\Gamma}^{s.p.}$ 
is about 28$\%$. For the relatively broad $s$-wave resonance 2$^-$, in which
$\Gamma_l/E_{R} \sim 0.2$, this discrepancy is about 13$\%$. For
the broad $s$-wave resonance 1$^-$ we have not succeeded to determine
${\cal R}_{\Gamma}^{s.p.}$ because in the single-particle potential model
its width is comparable to the resonance width.

\begin{table*}
\caption{
Ratios ${\cal R}_{\Gamma_0}^{MCM}$ and ${\cal R}_{\Gamma}^{MCM}$
obtained in the MCM in comparison with the prediction 
${\cal R}_{0}^{res}$ of the analytical formula (\ref{rnres})
and single-particle estimate ${\cal R}_{\Gamma}^{s.p.}$ (all given in
units of $\hbar c$) for the mirror pairs from the first column.
Also shown are spins and parities $J^{\pi}$ of the mirror states  
and orbital momenta $l$ of the relative motion in the resonance.
The results of the MCM calculations are averaged over two
NN potentials. For $^{13}$N, 
$\Gamma_{lS}^0$ obtained with a larger value of $a$ 
is used to calculate  ${\cal R}_{\Gamma_0}^{MCM}$.
 } 
\begin {center}
\begin{tabular}{ p{1.8 cm} p{0.8 cm} p{0.6 cm} p{3.4 cm} p{3.4 cm}
p{3.4 cm} p{3.4 cm}  }
\hline 
\\
 Mirror pair & $J^{\pi}$ &  $l$ & ${\cal R}_{\Gamma_0}^{MCM}$ &
${\cal R}_{0}^{res}$ & 
${\cal R}_{\Gamma}^{MCM}$ & ${\cal R}_{\Gamma}^{s.p.}$
 \\
\hline
$^8$Li-$^8$B & 1$^+$ &  1 & $(1.80\pm 0.04)\times 10^{-2}$ &
$(2.06\pm 0.04)\times 10^{-2}$ & $(1.73\pm 0.03)\times 10^{-2}$ & 
$(1.78\pm 0.01)\times 10^{-2}$ 
\\
   $^{12}$B-$^{12}$N & 2$^+$  & 1 & $(1.22\pm 0.02)\times 10^{-5}$ &
$(1.43\pm 0.01)\times 10^{-5}$ & $(1.20\pm 0.02)\times 10^{-5}$ & 
$1.22\times 10^{-5}$ 
\\
 $^{12}$B-$^{12}$N & 0$^+$  & 1 & $(2.07\pm 0.02)\times 10^{-2}$ &
$(2.51\pm 0.05)\times 10^{-2}$ & $2.01\times 10^{-2}$ & 
$(2.77\pm 0.14)\times 10^{-2}$ 
\\ 
 $^{12}$B-$^{12}$N & 2$^-$  & 0 & $ 2.79 \times 10^{-4}$ &
$(3.83\pm 0.31)\times 10^{-4}$ &$(2.23\pm 0.02)\times 10^{-4}$  
& $(2.55\pm 0.08)\times 10^{-4}$ 
\\     
  $^{12}$B-$^{12}$N & 1$^-$  & 0 & $(3.31\pm 0.09)\times 10^{-3}$ &
$(3.97\pm 0.56)\times 10^{-3}$ & $(2.98\pm 0.08)\times 10^{-3}$ &   
  \\  
$^{13}$C-$^{13}$N$^{2c}$ & $\frac{1}{2}^+$   & 0 & 
 $(7.13\pm 0.11)\times 10^{-5}$  & 
$(9.64\pm 0.81)\times 10^{-5}$ & $(5.59\pm 0.21)\times 10^{-5}$ & 
$(6.21 \pm 0.25)\times 10^{-5}$
\\  
$^{13}$C-$^{13}$N$^{4c}$  &  $\frac{1}{2}^+$  & 0 & 
$(7.49\pm 0.12)\times 10^{-5}$ &
$(9.64\pm 0.81)\times 10^{-5}$ & $(5.77\pm 0.22)\times 10^{-5}$ & 
$(6.21\pm 0.25)\times 10^{-5}$ 
\\                                               
$^{13}$C-$^{13}$N$^{2c}$ & $\frac{3}{2}^-$  & 1 &  
$(6.01\pm 0.09)\times 10^{-3}$ &
$(7.64\pm 0.06)\times 10^{-3}$ & $(5.96\pm 0.08)\times 10^{-3}$ & 
$(6.98\pm 0.22)\times 10^{-3}$ 
\\  
$^{13}$C-$^{13}$N$^{4c}$ & $\frac{3}{2}^-$ &   1 & 
$(6.39\pm 0.10)\times 10^{-3}$ &
$(7.64\pm 0.06)\times 10^{-3}$ & $(6.32\pm 0.14)\times 10^{-3}$ & 
$(6.98\pm 0.22)\times 10^{-3}$ 
\\      
$^{13}$C-$^{13}$N$^{2c}$ & $\frac{5}{2}^+$&   2 & 
$1.27 \times 10^{-2}$ &
$(1.43\pm 0.01)\times 10^{-2}$ & $(1.24\pm 0.01)\times 10^{-2}$ & 
$(1.37\pm 0.03)\times 10^{-2}$ 
\\  
$^{13}$C-$^{13}$N$^{4c}$ & $\frac{5}{2}^+$ &    2 & 
 $(1.32\pm 0.01)\times 10^{-2}$&
$(1.43\pm 0.01)\times 10^{-2}$ & $1.30\times 10^{-2}$ &
 $(1.37\pm 0.03)\times 10^{-2}$ 
\\ 
$^{23}$Ne-$^{23}$Al$^{a)}$ & $\frac{1}{2}^+$ & 0  & 
$(2.79\pm 0.10)\times 10^{-8}$ &
$(3.22\pm 0.13)\times 10^{-8}$ & $(2.76\pm 0.09)\times 10^{-8}$ & 
$(2.69\pm 0.02)\times 10^{-8}$ 
\\   
$^{23}$Ne-$^{23}$Al$^{b)}$ & $\frac{1}{2}^+$    & 0 & 
$(2.53\pm 0.02)\times 10^{-8}$ & $(3.22\pm 0.13)\times 10^{-8}$
 & $(2.44\pm 0.09)\times 10^{-8}$ & $(2.69\pm 0.02)\times 10^{-8}$ 
\\   
$^{26}$Mg-$^{27}$P$^{a)}$ &  $\frac{3}{2}^+$ & 2 & 
 $(1.75\pm 0.19)\times 10^{-11}$&
$2.21   \times 10^{-11}$ & $(1.76\pm 0.18)\times 10^{-11}$ & 
$(2.25\pm 0.10)\times 10^{-11}$ 
\\
$^{26}$Mg-$^{27}$P$^{b)}$ &  $\frac{3}{2}^+$  & 2 &
$(2.04\pm 0.02)\times 10^{-11}$ &
$2.21  \times 10^{-11}$ & $(2.04\pm 0.02)\times 10^{-11}$ & 
$(2.25\pm 0.10)\times 10^{-11}$ 
\\
\hline
\multicolumn{7}{l}{$^{2c}$ - two-cluster model} \\
\multicolumn{7}{l}{$^{4c}$ - four-cluster model }\\
\multicolumn{7}{l}{$^{a)}$ - multi-channel cluster model} \\
\multicolumn{7}{l}{$^{b)}$ - single-channel cluster model} \\
\end{tabular}
 
\end{center}
\label{table4}
\end{table*}

The neutron ANCs and r.m.s. radii for the  $\la^{12}$B$(2^-)|^{11}$B$\ra$ 
and $\la^{12}$B$(1^-)|^{11}$B$\ra$ overlap integrals
have been reported in Ref. \cite{Liu} where they have been determined
from the $^{11}$B(d,p)$^{12}$B reaction. 
The experimental values  of $C_1^2$  are $ 1.80 \pm 0.32$ fm$^{-1}$
and $0.88 \pm 0.15$ fm$^{-1}$ for the $2^-$ and $1^-$ respectively.
They are about 30 to 50$\%$ lower than the predictions of the MCM.

The experimental value of 
the r.m.s. radius for  2$^-$,
$\la r^2_{exp} \ra^{1/2}$ =
 $4.01 \pm 0.61$ fm from Ref. \cite{Liu}, is lower than the MCM
predictions $\la r^2 \ra^{1/2}$ = 4.99 and 4.78 fm,
while for the 1$^-$ state our predictions,  6.49 fm and 6.25 fm, agree with
the experimental value $\la r^2_{exp} \ra^{1/2}$ = $5.64 \pm 0.90$ fm
within the error bars.

Using the experimentally measured ANCs for $^{12}$B(2$^-$) and 
$^{12}$B(1$^-$) and the widths of the $^{12}$N(2$^-$) and
$^{12}$N(1$^-$) resonances from Ref. \cite{Aiz90}, we obtain
${\cal R}_{\Gamma}^{exp}$ = $(3.32\pm 0.98)\times 10^{-4}$ $\hbar c$ and
$(4.3\pm 2.2)\times 10^{-3}$ $\hbar c$ for the 2$^-$ and 1$^-$ states
respectively.  For the
2$^-$ resonance, the experimental value ${\cal R}_{\Gamma}^{exp}$ 
is larger than the theoretical ratio
$(2.23\pm 0.02)\times 10^{-4}$ $\hbar c$.
For the $1^-$ state,
the error bars of ${\cal R}_{\Gamma}^{exp}$ are too large to make conclusive
judgement about the agreement with theoretical calculations.

\subsubsection{$^{13}{\rm C} - ^{13}{\rm N}$}

We have used two different models  to describe the mirror pair 
$^{13}$N - $^{13}$C:
the multichannel two-cluster model $^{12}$C + n(p) 
of Ref. \cite{TBD97} and the multichannel four-cluster model
$\alpha+\alpha+\alpha$ + n(p) from Ref. \cite{Duf97}.
The difference between these two models
in predictions of ANCs, spectroscopic factors
and proton widths  reaches somethimes 30$\%$ or more
(see Tables II and III). 
In our previous work \cite{Tim05}, the
two- and four-cluster calculations gave even larger difference 
in mirror ANCs of ground states of the
mirror pair $^{13}$N - $^{13}$C.
As explained in Ref. \cite{Tim05}, 
such a large difference arises because the 
$\alpha+\alpha+\alpha$ model for the remnant nucleus 
$^{12}$C, used in the four-cluster calculations, contains only one type of
the permutational symmetry, namely, the one determined by 
the Young diagrams $[f]$ = [444]. 
One-center shell model of $^{12}$C, used in the
two-cluster calculations, contains all the other types of permutational
symmetry which may give singificant contributions to nuclear
properties associated with one nucleon removal.

The sensitivity of the ANCs squared and the proton widths 
on the NN potential choice is different for each 
mirror pair of excited states, however is does not exceed 30$\%$.
Dependence of the ratio  ${\cal R}_{\Gamma_0}^{MCM}$ or
${\cal R}_{\Gamma}^{MCM}$ on the NN potential is weaker, less than 5$\%$ for
the $\frac{3}{2}^-$ and $\frac{5}{2}^+$ states and 7$\%$ for the
$\frac{1}{2}^-$ state.

The calculated ratio  ${\cal R}_{\Gamma_0}^{MCM}$ is smaller than
the analytical estimate ${\cal R}_{0}^{res}$ by 8 to 22$\%$ for
narrow resonances $\frac{3}{2}^-$ and $\frac{5}{2}^+$ and and by
32$\%$ for the wide $s$-wave resonance $\frac{1}{2}^+$. The other ratio,
${\cal R}_{\Gamma}^{MCM}$, compares with the single-particle estimate
${\cal R}_{\Gamma}^{s.p.}$ more favourably, it is smaller
than ${\cal R}_{\Gamma}^{s.p.}$ by 6 to 15$\%$.

The neutron ANCs and r.m.s. radii for the  
$\la^{13}$C$(\frac{1}{2}^+)|^{12}$C$\ra$ 
and $\la^{13}$C$(\frac{5}{2}^+)|^{12}$C$\ra$  overlap integrals
have been reported in Ref. \cite{Liu} where they have been determined
from the $^{12}$C(d,p)$^{13}$C reaction. 
The experimental values  of $C^2$ are
$ 3.39 \pm 0.59$ fm$^{-1}$
and $0.023 \pm 0.003$ fm$^{-1}$ for the $\frac{1}{2}^+$ and 
$\frac{5}{2}^+$ respectively.
For $\frac{1}{2}^+$, the ANC squared $C^2$ = 
$ 3.65 \pm 0.34$  (statistical error) $\pm$ 0.35
(systematic error) fm$^{-1}$ has been independently measured
 in  Ref. \cite{Ima01} using the same reaction.
This value is slightly larger than the one from Ref.  \cite{Liu} but
agrees with it within the error bars. 
The  four-cluster MCM calculations with V2 potential
and  the two-cluster calculations with both  NN potentials
give for $C^2$ the values of 3.39, 3.46 and 3.41 fm$^{-1}$
that agree with the experimentally measured one. 
The four-cluster calculations with MN give slightly smaller ANC squared,
2.77 fm$^{-1}$.
As for the $\frac{5}{2}^+$ state, only the two-cluster calculations
with MN and the four-cluster calculations with V2,
$C^2$ = 0.0222 fm$^-1$ and  $C^2$ = 0.0213 fm$^-1$
respectively, agree with the experimentally determined value. 
The other two calculations either overestimate or underestimate it.

\begin{figure*}[t]
\centerline{\psfig{figure=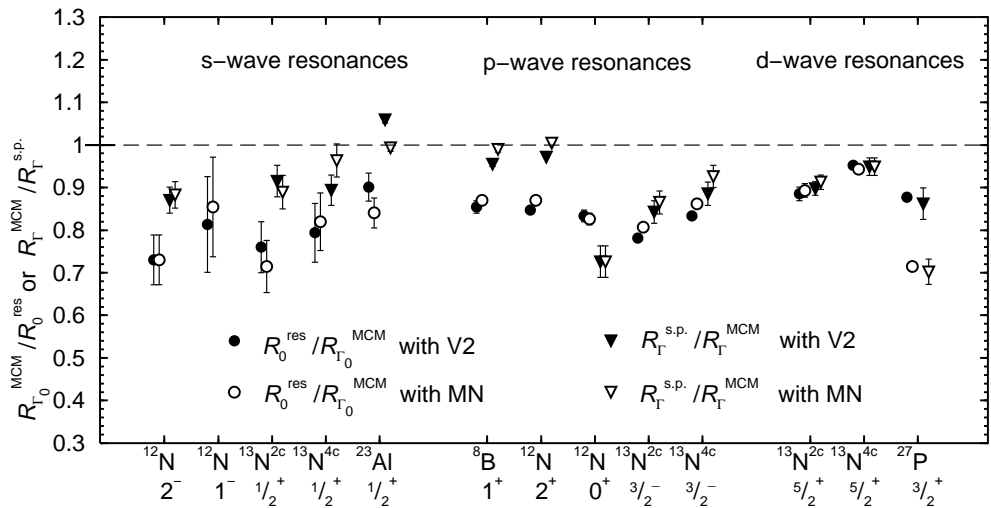,width=0.65\textwidth}
        }
\caption{Ratio between the MCM calculations
${\cal R}_{\Gamma_0}^{MCM}$
and the 
predictions ${\cal R}_0^{res}$
of the analytical formula (\ref{rnres}) and ratio between 
${\cal R}_{\Gamma}^{MCM}$ and the single-particle estimate
${\cal R}_{\Gamma}^{s.p.}$. Both are calculated 
with two NN potentials, V2  and MN.
Core excitations are taken into account and 
charge symmetry of NN interactions is broken. 
 Both four-cluster ($4c$) and two-cluster ($2c$) 
calculations for $^{13}$N are shown. 
}
\end{figure*}

The experimental value of 
the r.m.s. radii,
$\la r^2_{exp} \ra^{1/2}$ = $5.04 \pm 0.75$ fm and 
 $\la r^2_{exp} \ra^{1/2}$ = $3.68 \pm 0.40$ fm
 for  the  $\frac{1}{2}^+$ and $\frac{5}{2}^+$ states in $^{13}$C
 respectively, have been reported in Ref. \cite{Liu}. They
agree with the MCM predictions obtained both in two- and four-cluster model
with V2 and MN potentials within the error bars.


Using  the  ANCs measured in Ref. \cite{Liu,Ima01}
for the $\frac{1}{2}^+$ and $\frac{5}{2}^+$ states
in $^{13}$C and the
experimental proton widths for their mirror analogs
available in  Ref. \cite{Ajz91},
we can construct the  ratio ${\cal R}_{\Gamma}^{exp}
= \Gamma_p^{exp}/(C^{exp}_n)^2$.
For $\frac{1}{2}^+$, this ratio is
$(4.74 \pm 0.94)\times 10^{-5}$ $\hbar c$  or
$(4.40 \pm 0.94)\times 10^{-5}$ $\hbar c$ depending on
whether  the neutron ANC used is taken from Ref. \cite{Liu}
or \cite{Ima01}.
The MCM ratios 
${\cal R}_{\Gamma}^{MCM}$  of $(5.77\pm 0.22)\times 10^{-5}$ $\hbar c$
and $(5.59\pm 0.21)\times 10^{-5}$ $\hbar c$,
obtained in four- and two-cluster calculations respectively,
agree with the first  value for ${\cal R}_{\Gamma}^{exp}$  
within the error bars but
are larger than the second value.
As for the $\frac{5}{2}^+$ state, all the MCM calculations of the ratio 
${\cal R}_{\Gamma}^{MCM}$  agree within the error bars with 
${\cal R}_{\Gamma}^{exp}$  =
$(1.04 \pm 0.29)\times 10^{-2}$ $\hbar c$,
obtained using the  experimental proton width of from 
\cite{Ajz91}
and the neutron ANC from \cite{Liu}.

\subsubsection{$^{23}{\rm Al} - ^{23}{\rm Ne}$}

The calculated ANCs  
and  spectroscopic factors for $^{23}$Ne($\frac{1}{2}^+$)
as well as  the proton widths of 
$^{23}$Al($\frac{1}{2}^+$) are strongly influenced by the excitations
in the $^{22}$Ne and $^{22}$Mg cores.  
Including the core excitations decrease 
$C_l^2$, $S_l$ and $\Gamma_p$
by 4 to 7 times. Dependence of these values on the 
NN potential choice is much weaker, $\sim 12\%$ in the single-channel
calculations and $\sim 50\%$ in the multi-channel calculations.
The dependence of the ratio ${\cal R}_{\Gamma_0}^{MCM}$ on the NN
potential is even weaker, 2$\%$ for single-channel calculations
and 7$\%$ for multi-channel calculations. For ${\cal R}_{\Gamma}^{MCM}$,
this dependence is about 7$\%$. The core excitations lead to a 10$\%$
increase in ${\cal R}_{\Gamma_0}^{MCM}$ and a 13$\%$ increase in
${\cal R}_{\Gamma}^{MCM}$.

The MCM ratio ${\cal R}_{\Gamma_0}^{MCM}$ is about 20$\%$ lower than
the predictions ${\cal R}_{0}^{res}$  of the analytical formula 
(\ref{rnres}). The other ratio, ${\cal R}_{\Gamma}^{MCM}$, is very
close to the single-particle estimate ${\cal R}_{\Gamma}^{s.p.}$,
if it is calculated with core excitations taken into account.
Without taking core excitations into account, it is 10$\%$ smaller
than  ${\cal R}_{\Gamma}^{s.p.}$.

\subsubsection{$^{27}{\rm P} - ^{27}{\rm Mg}$}

As in the case of $^{23}$Ne - $^{23}$Al, the
MCM predictions strongly depend on whether the core excitations are taken
into account or not. Without   core excitations, the dependence of  
ANCs squared and proton widths on the NN potential choice is relatively weak,
about 30$\%$. When the core excitations are included, the spectroscopic
factor for the $\la ^{27}$Mg$(\frac{3}{2}^+)|^{26}$Mg(g.s.)$\ra$ overlap 
drops by 30$\%$ and 70$\%$ for the V2 and MN potential respectively.
The $C^2$ and $\Gamma_p$ 
values decrease approximately by the same amount too.
Their dependence on the NN potential becomes much stronger, about 3 times.

The dependence of the ratio  ${\cal R}_{\Gamma_0}^{MCM}$ and
${\cal R}_{\Gamma}^{MCM}$ on the NN potential in single-channel calculations
is about 2$\%$, but in the multichannel calculations this
dependence increases  up to 20$\%$. In both cases, $C^2$ and $\Gamma_p$
are more sensitive to the NN potential than their ratio.

The single-channel calculations of ${\cal R}_{\Gamma_0}^{MCM}$ 
and ${\cal R}_{\Gamma}^{MCM}$ are
close  to ${\cal R}_{0}^{res}$ and ${\cal R}_{\Gamma}^{s.p.}$.
The core excitations decrease ${\cal R}_{\Gamma_0}^{MCM}$ 
and ${\cal R}_{\Gamma}^{MCM}$ by about 15$\%$.


\section{Mirror symmetry in spectroscopic factors}

Spectroscopic factors are often used to predict the widths of proton 
resonances in approaches of a shell model type. Two questions
concerning  this procedure arise: how reliable are the concepts
of spectroscopic factors for unbound states and is there any mirror symmetry
between such spectroscopic factors and their mirror analogs. In this section,
we briefly address these questions from the point of view of the MCM.

The spectrosocpic factor $S_{lj}$ for a particle-bound state is defined as
\beq
S_{lj} = A \int_0^{\infty} dr \, r^2 (I_{lj}(r))^2,
\eeqn{sf}
where $I_{lj}(r)$ is a radial part of the overlap integral
between the wave functions of nuclei $A$ and $A-1$.
For unbound states, the contribution from the oscillating tail
of this integral will lead to a divergent result for  $S_{lj}$.
In order to use the concept of the spectroscopic factor for unbound states,
its definition should be modified. In the present paper, we define it
as  the norm of the overlap integral 
$I^{BSA}(r)$ between
the wave function of nucleus $A$ obtained in the bound state
approximation  and the wave function of nucleus $A-1$.

The calculated proton overlap integrals decrease very slowly inside
the channel radius $a$. For $l \neq 0$ resonances in
the 0$p$-shell nuclei, the function $rI^{BSA}(r)$
typically decrease from its maximum value only by a factor of two or three.
It decreases even slowlier for the $s$-wave resonance $^{13}$N($\frac{1}{2}^+$).
We would like to note here  that such a slow decrease is absent in
shell model approaches.
For very narrow $sd$-shell
resonances $^{23}$Al($\frac{1}{2}^+$) and $^{27}$P($\frac{3}{2}^+$) 
with the widths
of $\sim 10^{-2}$ and   $\sim 10^{-6}$ keV the strong Coulomb barrier
traps their wave functions inside the channel radius so that
$rI^{BSA}(r)$ decreases from its maximum value by an order of magnitude.
This resembles more   the conventional  bound state behaviour of
the wave functions so that shell model description for these states
should be more adequate.

The calculated spectroscopic factors  depend  on the choice 
the channel radius. We have performed the calculations for $^{13}$N
with two channel radii shown in Table III. Also, for $^{27}$P, 
the calculations have been performed with $a$ = 9.0 and 9.5 fm.
The change in spectroscopic factors is less than 4$\%$ in all these cases.
The  ratio $S_p/S_n$
between the mirror spectroscopic factors, 
calculated with larger value of $a$ when available, is presented in Fig.5.
This figure shows significant symmetry breaking in spectroscopic factors,
which $\sim 20\%$ 
for $s$-wave resonances and in $^{13}$N($\frac{3}{2}^-$),
$\sim$ 5-8$\%$ for $^{8}$B, $^{12}$N(2$^+$), $^{12}$N(0$^+$) and
$^{13}$N($\frac{5}{2}^+$), and for $^{27}$P the charge symmetry breaking
depends of the NN potential used in caluclations: 
it is $\sim$ 4$\%$ for V2 and $\sim$ 25$\%$ for MN. This is larger than
the 3-9$\%$ charge-symmetry breaking  in spectroscopic factors for bound-bound
mirror pairs obtained in Ref. \cite{Tim05}.

\begin{figure}[t]
\centerline{\psfig{figure=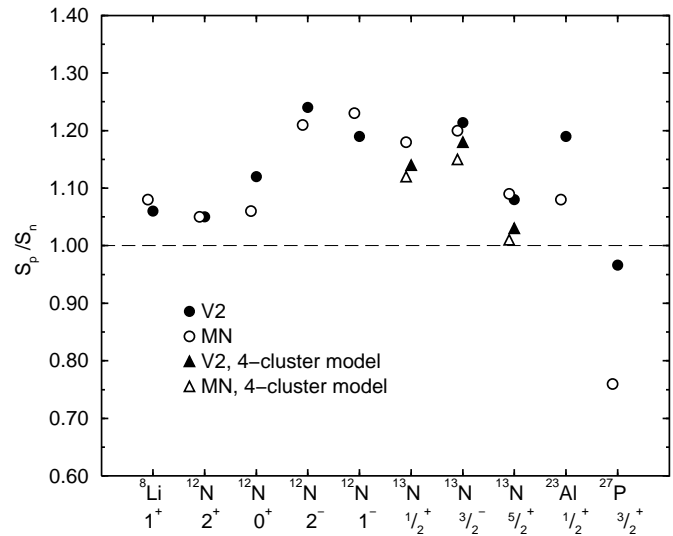,width=0.49\textwidth}
       }
\caption{Ratio between the proton  and mirror neutron spectroscopic factors
calculated within the MCM. For the $^{13}$N - $^{13}$C mirror pair, 
the circles represent the calculations obtained within the two-cluster MCM.
}
\end{figure}

\section{Summary and conclusions}

Due to the charge symmetry of NN interactions
the width of a narrow proton resonance is related to the
ANC of its  mirror neutron bound state.
This relation (the ratio ${\cal R}_{\Gamma}$)
can be approximated by a simple  
analytical formula  (\ref{rnres}) derived in Ref. \cite{Tim03}.
In this paper, we have clarified that
the width that enters formula (\ref{rnres})
has the meaning of a residue in the R-matrix pole.
In most cases, however, the widths that have been
determined from experimental data are associated
with Breit-Wigner shapes of the resonant cross sections.
The MCM calculations, performed in this paper for a range
of nuclei, have shown that these two widths can differ
by less than 7$\%$ for narrow resonances and up to 30$\%$
for broad $s$-wave resonances. Since the most
interesting applications of ${\cal R}_{\Gamma}$,
for example, predictions of resonant proton capture rates
using mirror neutron ANCs,
require a knowledge of ``observed" widths,
investigation of a link between such ``observed" widths
with mirror neutron ANCs is also important.

In the present paper we have studied the ratio ${\cal R}_{\Gamma}$
within the MCM.
The calculated  MCM proton widths and neutron ANCs 
are very sensitive to the
details of the model and to the NN interaction choice, however, the ratios
${\cal R}_{\Gamma_0}^{MCM}$ and
${\cal R}_{\Gamma}^{MCM}$, based on two different definitions of widths, are
in most cases  almost model-independent.
Comparison of the ratio ${\cal R}_{\Gamma_0}^{MCM}$,associated
with the residue in the R-matrix pole,
with the predictions ${\cal R}_{0}^{res}$ of Eq. (\ref{rnres}) has confirmed
the general trend in its behaviour given by this formula. 
The difference
between  ${\cal R}_{\Gamma_0}^{MCM}$
and ${\cal R}_{0}^{res}$ are on average
about 20$\%$ (see Fig.4) and does not exceed 30$\%$ for the
range of nuclei considered here. 
This difference is   larger than the average 6$\%$ divergence
between the  ratio  of 
mirror ANCs for bound-bound mirror pairs obtained in the MCM
for the same nuclei and its analytical estimate \cite{Tim05}.
The other ratio, ${\cal R}_{\Gamma}^{MCM}$, associated with
``observed widths", is close to the estimate
${\cal R}_{\Gamma}^{s.p.}$, made in the single-particle model on the
assumptions that  mirror single-particle potential well 
are exactly the same. The average deviation between ${\cal R}_{\Gamma}^{MCM}$
and ${\cal R}_{\Gamma}^{s.p.}$ is about 10$\%$ and does not exceed 30$\%$
(see Fig.4). 
 The large deviation, about 30$\%$, obtained for the $^{27}$Mg - $^{27}$P
mirror pair with the MN potential, can be explained by the restricted 
model space used to generate the wave functions of the $^{26}$Mg and
$^{26}$Si cores.
Let us note that ${\cal R}_{\Gamma}^{MCM}$
is close to  ${\cal R}_{\Gamma}^{s.p.}$ even for broad $s$-wave resonances.

For few resonances, $^8$B(1$^+$), $^{12}$N(2$^-$,1$^-$) and
$^{13}$N($\frac{1}{2}^+,\frac{5}{2}^+$), 
the comparison of the ratio ${\cal R}_{\Gamma}^{MCM}$
with the ratio ${\cal R}_{\Gamma}^{exp}$ 
between experimentally measured widths and mirror neutron
ANCs squared has been possible. We have observed
agreement within the error bars between ${\cal R}_{\Gamma}^{MCM}$
and ${\cal R}_{\Gamma}^{exp}$ for
$^8$B(1$^+$), when ${\cal R}_{\Gamma}^{exp}$ has been constructed
using new value for $\Gamma_p$ from Ref. \cite{Ang03}. Also,
the agreement has been achieved
for $^{12}$N(1$^-$), $^{13}$N($\frac{5}{2}^+$)
and for $^{13}$N($\frac{1}{2}^+$), when ${\cal R}_{\Gamma}^{exp}$ 
is calculated using neutron ANC from Ref. \cite{Liu}.
For $^8$B(1$^+$), the ratio ${\cal R}_{\Gamma}^{exp}$ 
calculated with old value of $\Gamma_p$ from Ref. \cite{Aiz88}
is larger than
${\cal R}_{\Gamma}^{MCM}$. ${\cal R}_{\Gamma}^{exp}$ is also larger
than ${\cal R}_{\Gamma}^{MCM}$ for $^{12}$N(2$^-$).
For $^{13}$N($\frac{1}{2}^+$), ${\cal R}_{\Gamma}^{exp}$ 
constructed with neutron ANC from \cite{Ima01} is
smaller than ${\cal R}_{\Gamma}^{MCM}$.
The disagreement between ${\cal R}_{\Gamma}^{MCM}$
and ${\cal R}_{\Gamma}^{exp}$ indicates that, in the first
instance,  it is  necessary to   remeasure  neutron ANCs. 
Unlike ${\cal R}_{\Gamma}^{MCM}$,
which does not depend strongly on model assumptions,
${\cal R}_{\Gamma}^{exp}$ is constructed using ANCs 
that are determined via theoretical analysis of
some neutron removal reactions. The systematical errors
of such an analysis, for example, due to uncertainty in optical
potential choice or because of influence of breakup effects,
may be as high as 30$\%$. Uncertanties in proton widths
are usually smaller, but as the case of $^8$B(1$^+$)
has shown, they can noticeably influence
${\cal R}_{\Gamma}^{exp}$.

The knowledge of ${\cal R}_{\Gamma}$ can be used to predict
proton widths if the mirror neutron ANCs are known. It can have 
important astrophysical application to predict the  proton capture rates 
via the resonances for which $\Gamma_p < \Gamma_{\gamma}$.
Our calculations have shown that for each resonance the different model 
assumptions and NN potentials give the deviation from the average
value of ${\cal R}_{\Gamma}^{MCM}$ no more than 10$\%$. This means that
if neutron ANCs squared are measured with an accuracy of 10-20$\%$, then
the proton widths can be determined with the accuracy of 10-30$\%$.
This is better than the use of unreliable concept of spectroscopic factors in
continuum and analysis of stripping  reactions to unbound states
for the same purposes.

\section*{Acknowledgements}
T.N.K. is grateful to Professors R.C. Johnson, D. Baye and I.J. Thompson
for useful discussions.
Support from the UK EPSRC via grant GR/T28577 is greatfully acknowledged.

\end{document}